\documentstyle [pra,aps,preprint]{revtex}
\begin{document}
\draft
\title{Towards a quantum Hall effect for atoms using 
electric fields}
\author{Marie Ericsson$^{1}$\footnote{E-mail: 
marie.ericsson@kvac.uu.se} and
Erik Sj\"{o}qvist$^{1}$\footnote{E-mail: 
erik.sjoqvist@kvac.uu.se}}
\address{$^{1}$Department of Quantum Chemistry, Uppsala 
University, Box 518, Se-751 20 Uppsala, Sweden}
\maketitle
\begin{abstract}
An atomic analogue of Landau quantization based on the 
Aharonov-Casher (AC) interaction is developed. The effect 
provides a first step towards an atomic quantum Hall system 
using electric fields, which may be realized in a Bose-Einstein 
condensate. 
\end{abstract}
\pacs{PACS numbers: 03.65.-w, 03.75.Fi, 11.30.Pb, 73.43.-f}  
\section{Introduction}
Paredes {\it et al.} \cite{paredes01} have recently proved 
the existence of anyonic excitations in rotating Bose-Einstein 
condensates, demonstrating a method to test exotic particle 
statistics. Their analysis was based on the analogy between 
a condensate in a rotating trap and a system of interacting 
electrons in a uniform static magnetic field. 

Motivated by this result, we provide in this work a first 
step towards another atomic quantum Hall analogy that could 
be of interest for the physics of Bose-Einstein condensates. 
The idea is based on the Aharonov-Casher (AC) effect 
\cite{aharonov84} (see also \cite{anandan82}) in which 
atoms may interact with an electric field via a nonvanishing 
magnetic moment. This interaction coincides formally in the 
nonrelativistic limit with that of minimal coupling, where 
the AC vector potential is determined by the electric field 
and the direction of the magnetic dipole. We demonstrate 
the existence of a certain field-dipole configuration in 
which an atomic analogue of the standard Landau effect 
\cite{landau30} occurs. This result opens up the possibility 
for an atomic realization of the quantum Hall effect using 
electric fields. It should be noted that the AC interaction 
in Bose-Einstein condenstates has previously been discussed 
\cite{petrosyan99}, but in the different context related 
to vortex formation. 
 
In the next section, we briefly outline the standard 
Landau theory for a charged particle moving in a uniform 
magnetic field. The precise conditions under which the AC 
analogue of the Landau effect occurs are stated in section 
III. Under fulfillment of these conditions the corresponding 
theory is developed in detail. The relation to the AC duality 
\cite{aharonov84} as well as aspects of gauge and supersymmetry 
are delineated in section IV. These aspects illuminates an 
additional richness in the physics of the present Landau 
effect compared to that of the standard one. The paper ends 
with concluding remarks.  

\section{Standard Landau theory}
Consider a particle with charge $q$ moving in a plane 
perpendicular to a uniform magnetic field ${\bf B} = 
B{\bf z}$, say. This system is the basic constituent 
of the quantum Hall effect and is described by Hamiltonian 
operator (SI units are used throughout this paper) 
\begin{equation}
H = \frac{{\bf {\Pi}}^{2}}{2m} = 
\frac{1}{2m} (-i\hbar \nabla - q{\bf A})^{2} , 
\end{equation}
where $m$ is the mass of the particle, $2\pi\hbar$ Planck's 
constant, and ${\bf A}$ is the vector potential fulfilling 
${\bf B} = \nabla \times {\bf A}$. The eigenvalues of $H$ 
are computed using standard ladder operator technique 
based on the canonical structure of the commutator 
\begin{equation}
[\Pi_{x},\Pi_{y}] = i\hbar m \omega = 
i \sigma \hbar^{2} / \ l^{2} .
\end{equation}
Here $\omega = qB/m = \sigma |qB|/m$ is the cyclotron  
frequency for which the sign $\sigma$ describes the revolution  
direction of the corresponding classical motion. The natural 
unit of length in the quantum Hall regime is the magnetic 
length $l = \sqrt{\hbar /|qB|}$. Next we introduce the 
annihilation and creation operators  
\begin{eqnarray}
a = \frac{1}{\sqrt{2m\hbar | \omega |}} 
(\Pi_{x} + i\sigma\Pi_{y}) 
\nonumber \\
a^{\dagger} = \frac{1}{\sqrt{2m\hbar | \omega |}} 
(\Pi_{x} - i\sigma\Pi_{y}) 
\end{eqnarray}
that fulfill the commutation relation $[a,a^{\dagger}]=1$.  
In terms of these, we may write the Hamiltonian operator as 
\begin{equation}
H = (a^{\dagger}a + \frac{1}{2}) \hbar | \omega | + 
\frac{p^{2}_{z}}{2m}. 
\end{equation}
Thus the motion in the $x-y$ plane has been transformed into 
a one-dimensional harmonic oscillator accompanying free motion 
in the $z$ direction. It follows that the energy eigenvalues are 
\begin{eqnarray}
E_{\nu,k_{z}} & = & 
\Big(\nu+\frac{1}{2} \Big) \hbar | \omega | + 
\frac{\hbar^{2}k_{z}^{2}}{2m}, 
\nonumber \\ 
\nu & = & 0,1,2,..., 
\nonumber \\ 
k_{z} & & {\text{real-valued.}} 
\label{eq:landauenergies}
\end{eqnarray}
Note that these eigenvalues are independent of both 
the revolution direction and the orbit center of the 
corresponding classical motion. The latter independence 
is related to the fact that the above energy eigenvalues 
are degenerate. This degeneracy is revealed by labeling 
the corresponding eigenfunctions with the eigenvalue $\eta$ 
of the quantum orbit center operator \cite{macdonald94}. 
Such a degenerate set $\{ \psi_{\eta,\nu,k_{z}} \}_{\eta}$ 
defines a Landau level \cite{landau30}.  

\section{AC analogue of Landau theory}
In the nonrelativistic limit one may describe the interaction 
between an atom with nonvanishing magnetic moment $\mu$ and 
an electric field ${\bf E}$ by the AC Hamiltonian operator 
(neglecting terms of $O({\bf E}^{2})$) \cite{anandan89}
\begin{equation}
H=\frac{{\bf {\Pi}}^{2}}{2m}+
\frac{\mu \hbar}{2 m c^{2}}{\bf \nabla}\cdot{\bf E}.
\label{eq:hamiltonian}
\end{equation}
Here ${\bf \Pi} = -i \hbar \nabla - \mu c^{-2}({\bf n}\times 
{\bf E})$ is the kinematic momentum with ${\bf n}$ the direction 
of the magnetic dipole moment $\mbox{\boldmath$\mu$}$, $m$ is 
the mass of the particle, and $c$ is the speed of light. This 
defines the AC vector potential 
\begin{equation}
{\bf A}_{AC} = c^{-2} ({\bf n} \times {\bf E}) , 
\end{equation}
the associated field strength
\begin{equation}
{\bf B}_{AC} = c^{-2} \nabla \times 
({\bf n} \times {\bf E}) , 
\label{eq:beff}
\end{equation}
and the coupling strength $\mu$. It follows that 
\begin{equation}
[\Pi_{k},\Pi_{l}] = 
i\hbar\mu\epsilon_{klm} 
({\bf B}_{AC})_{m},
\label{eq:comm1}
\end{equation}
where $\epsilon_{klm}$ is the Levi-Civita symbol with 
$k,l,m$ running over $x,y,z$ and the summation convention 
is used. 

The precise conditions on the field-dipole configuration 
under which the AC analogue of the Landau effect occurs are: 
\begin{enumerate}
\item[(i)] Condition for vanishing torque on the dipole: 
${\bf n} \times (\langle {\bf {\Pi}} \rangle \times {\bf E})=0$, 
where $\langle \cdot \rangle$ denotes expectation value. 
\item[(ii)] Conditions for electrostatics: 
$\partial_{t} {\bf E} = 0$ and $\nabla \times{\bf E}=0$. 
\item[(iii)] ${\bf B}_{AC}$ uniform. 
\end{enumerate}

Condition (i) follows from the fact that a dipole moving with 
velocity $\langle {\bf {\Pi}} \rangle$ sees an effective magnetic 
field ${\bf B}_{\text{eff}} \propto \langle {\bf {\Pi}} \rangle 
\times {\bf E}$ in its own reference frame and that this magnetic 
field produces a torque $\dot{{\bf n}} \propto {\bf n} \times 
{\bf B}_{\text{eff}}$. 

For ${\bf n} = (0,0,1)$, conditions (i) and (ii) are fulfilled 
if  ${\bf E} = (E_{x} (x,y),E_{y} (x,y),0)$ with $\partial_{x} 
E_{y} -\partial_{y} E_{x} = 0$ and the atom moves in the 
$x-y$ plane so that $\langle \Pi_{z} \rangle$ vanishes. 
Using the vector identity ${\bf B}_{AC} = {\bf n} (\nabla 
\cdot {\bf E}) + ({\bf n} \cdot \nabla ) {\bf E}$, the 
second term vanishes and condition (iii) reduces to Gauss' 
law $\nabla \cdot {\bf E} = \rho_{0}/\epsilon_{0}$, where 
$\rho_{0}$ is a nonvanishing uniform volume charge density 
and $\epsilon_{0}$ is the electric vacuum permittivity.  

With the Landau conditions fulfilled and the above choice 
of ${\bf n}$ and ${\bf E}$, we may proceed as in the 
preceding section to work out energy eigenstates using 
standard ladder operator technique. The kinetic part of the 
Hamiltonian operator is taken care of by noting that the 
only nonvanishing commutator in Eq.~(\ref{eq:comm1}) is 
\begin{equation}
[\Pi_{x},\Pi_{y}] = i\hbar m \omega_{AC} = 
i \sigma \hbar^{2} / \ l_{AC}^{2} 
\label{eq:comm2}
\end{equation}
with the cyclotron frequency  
\begin{equation}
\omega_{AC}=\frac{\mu \rho_{0}}{m c^{2}\epsilon_{0}} = 
\frac{\sigma |\mu \rho_{0}| }{m c^{2}\epsilon_{0}} . 
\label{eq:larmor1}
\end{equation}
Again $\sigma = \pm 1$ labels the revolution direction of the 
corresponding classical motion. The natural unit of length 
in the AC case is $l_{AC} = \sqrt{\hbar c^{2}\epsilon_{0} / 
|\mu\rho_{0}|}$. Next we define the annihilation 
and creation operators 
\begin{eqnarray}
a_{AC} & = & \frac{1}{\sqrt{2m\hbar |\omega_{AC}|}} 
(\Pi_{x}+i\sigma\Pi_{y})
\nonumber \\
a_{AC}^{\dagger} & = & \frac{1}{\sqrt{2m\hbar |\omega_{AC}|}} 
(\Pi_{x}-i\sigma\Pi_{y}) 
\label{eq:crean}
\end{eqnarray}
that fulfill the commutation relation $[a_{AC},a_{AC}^{\dagger}]=1$. 
Inserting Eq.~(\ref{eq:crean}) into Eq.~(\ref{eq:hamiltonian}) we 
obtain 
\begin{equation}
H=\left( a_{AC}^{\dagger}a_{AC}+\frac{1}{2}
(1+\sigma) \right)\hbar |\omega_{AC}|+\frac{p_{z}^{2}}{2m} , 
\end{equation}
where we have used that 
\begin{equation}
\frac{\mu \hbar}{2 m c^{2}}{\bf \nabla}\cdot
{\bf E}=\frac{1}{2}\hbar \omega_{AC} . 
\label{eq:zero}
\end{equation}
With the constraint $\langle \Pi_{z} \rangle  = \langle p_{z} 
\rangle =0$ imposed by condition (i) we obtain the energy 
eigenvalues  
\begin{eqnarray}
E_{\nu}^{(\sigma)} & = &  
\left( \nu+\frac{1}{2}(1+\sigma)\right)\hbar |\omega_{AC}|, 
\nonumber \\ 
\nu & = & 0,1,2...
\label{eq:landau1}
\end{eqnarray}
These energies are independent of the classical orbit center 
but they depend on the revolution direction. The $\sigma$ 
dependent set of degenerate states $\{ \psi_{\eta,\nu}^{(\sigma)} 
\}_{\eta}$ defines the AC analogue of a Landau level ($\eta$ is 
again the eigenvalue of the quantum orbit center operator). Note 
that the condition for vanishing torque on the dipole puts an 
additional constraint on the stationary states 
$\psi_{\eta,\nu}^{(\sigma)}$, viz. that they must fulfill 
$\partial_{z} \psi_{\eta,\nu}^{(\sigma)} = 0$. 

\section{Physical interpretation of the AC analogue}
\subsection{Relation to the AC duality}
The physical origin of the AC analogue may be understood 
from the standard Landau effect using the AC duality 
\cite{aharonov84}
\begin{eqnarray}
q\Phi & \leftrightarrow & 
\frac{\mu \lambda}{c^{2}\epsilon_{0}} ,
\label{eq:duality}
\end{eqnarray}  
where $\Phi$ is a magnetic flux and $\lambda$ is a 
uniform linear charge density in the direction of the 
magnetic dipole. Consider the separation of the energies 
in Eq.~(\ref{eq:landauenergies}) for fixed $k_{z}$ 
\begin{equation}
\Delta E = \hbar \frac{|qB|}{m} .
\end{equation}
The magnetic field can be expressed in terms of its flux 
$\Phi$ as $B=\Phi /S$, $S$ being the area (perpendicular 
to the magnetic field) through which the flux is measured. 
This leads to
\begin{equation}
\Delta E = \hbar \frac{|q\Phi /S|}{m}.
\label{eq:separation}
\end{equation}
Thus, according to Eq.~(\ref{eq:duality}) the AC dual to 
$\Delta E$ is  
\begin{equation}
\Delta E_{AC} = \hbar \frac{|\mu\lambda /S|}{mc^2\epsilon_{0}} =
\hbar \frac{|\mu \rho_{0}|}{mc^2\epsilon_{0}},
\label{eq:reciprocalseparation}
\end{equation}
which is the Landau level separation in the AC case for fixed 
$\sigma$. We have used that $\rho_{0} = \lambda /S$ is a 
uniform volume charge density with the direction of $\lambda$ 
perpendicular to $S$.

\subsection{Gauge symmetries}
In the case of a charged particle interacting with an 
electromagnetic field, two distinct experimental set ups 
cannot be related by a gauge transformation. This is not 
so for the AC interaction, where there exist different 
choices of physical set ups associated with the same physical 
effect. The basic reason for this fact is that the AC vector 
potential is directly linked to physical quantities, viz. 
the electric field ${\bf E}$ and the direction of the 
magnetic dipole ${\bf n}$, in such a way that two different 
pairs $({\bf n},{\bf E})$ and $({\bf n}',{\bf E}')$ may 
yield the same ${\bf B}_{AC}$.   

To explore this additional gauge symmetry in more detail, let 
us consider the AC vector potential  ${\bf A}_{AC} = c^{-2} 
(-E_{y},E_{x},0)$ for ${\bf n}=(0,0,1)$ and ${\bf E} = 
(E_{x} (x,y),E_{y} (x,y),0)$. Any electric field 
${\bf E}'$ related to this ${\bf E}$ by
\begin{eqnarray}
E_{x}' & = & E_{x}+\partial_{y}\chi \nonumber \\
E_{y}' & = & E_{y}-\partial_{x}\chi 
\label{eq:electricgauge}
\end{eqnarray}
defines the AC vector potential 
\begin{equation}
{\bf A}'_{AC} = {\bf A}_{AC} + \nabla \chi  
\label{eq:vectorgauge}
\end{equation}
so that ${\bf B}_{AC}' = {\bf B}_{AC}$. In particular, with 
uniform ${\bf B}_{AC}$ the Landau conditions are fulfilled 
for the atom moving in the $x-y$ plane also for ${\bf E}'$ 
in Eq.~(\ref{eq:electricgauge}), if $E_{z}' = 0$, 
$\chi = \chi (x,y)$, and $\nabla^{2} \chi = 0$.  

In the Landau case with ${\bf n}=(0,0,1)$, it is in this 
context instructive to consider a uniform volume charge 
density $\rho_{0}$ confined to regions with differently 
shaped $x-y$ cross-sections. Solving for a cylindrical 
shape yields the electric field ${\bf E} = \rho_{0} / 
(2\epsilon_{0}) \ (x,y,0)$ in the interior of the cylinder. 
Here $x$ and $y$ are measured relative the symmetry axis of 
the cylinder, which we take to be the $z$ axis.
The corresponding vector potential becomes ${\bf A}_{AC} = 
\rho_{0} / (2c^{2}\epsilon_{0}) \ (-y,x,0)$. This is the AC 
analogue of symmetric gauge. On the other hand, within a 
uniformly charged plate of finite width in the $x$ direction, 
but infinite extension in the $y$ and $z$ directions, the 
electric field takes the form ${\bf E}=\rho_{0}/\epsilon_{0} 
\ (x,0,0)$ with $x$ measured relative one of the surfaces 
of the plate. The corresponding vector potential is 
${\bf A}_{AC} = \rho_{0}/(c^{2}\epsilon_{0}) \ (0,x,0)$. 
This is the AC analogue of Landau gauge. These two 
configurations yield identical AC Landau level energies 
and are related by the gauge function 
$\chi = \rho_{0} / (2\epsilon_{0}c^{2}) \ xy$. Other 
choices of gauge may be obtained by further changes 
of the shape of the $x-y$ cross-section. 

\subsection{Supersymmetry} 
The dependence of the Landau energies on the revolution 
direction in the AC case, divides the set of Landau 
levels into two classes, each labeled by the value of 
$\sigma$. As we show now, this may be understood in terms 
of supersymmetry \cite{roy93}. 

We start by reinterpreting $\sigma$ as the eigenvalue of 
an operator $\tau$ and introduce 
the supercharge 
\begin{equation}
Q = a_{AC} f^{\dagger} , 
\end{equation}
where $[f,f^{\dagger}] = \tau$, $ff^{\dagger}+f^{\dagger}f=1$, 
and $ff=f^{\dagger}f^{\dagger}=0$. This yields 
\begin{eqnarray}
H / (\hbar \omega_{AC} ) & = & QQ^{\dagger} + Q^{\dagger}Q 
\nonumber \\ 
 & = & a_{AC}^{\dagger} a_{AC} + \frac{1}{2} 
\Big( 1 -  [f,f^{\dagger}] \Big) .  
\end{eqnarray} 
Next we introduce the boson and fermion number operators 
\begin{eqnarray}
N_{B} & = & a_{AC}^{\dagger} a_{AC}  
\nonumber \\ 
N_{F} & = & \frac{1}{2} \Big( 1-\tau \Big) = 
\frac{1}{2} \Big( 1-[f,f^{\dagger}] \Big) . 
\end{eqnarray}
and the Fock space $\{ |n_{AC},n_{F} \rangle \}$ defined 
by the action of the operators $a$, $a^{\dagger}$,$f$, and 
$f^{\dagger}$:
\begin{eqnarray} 
a_{AC} |n_{B},n_{F} \rangle & = & 
\sqrt{n_{B}} |n_{B}-1,n_{F} \rangle 
\nonumber \\ 
a_{AC}^{\dagger} |n_{B},n_{F} \rangle & = & 
\sqrt{n_{B}+1} |n_{B}+1,n_{F} \rangle 
\nonumber \\ 
f |n_{B},n_{F} \rangle & = & 
|n_{B},n_{F}-1 \rangle 
\nonumber \\ 
f^{\dagger} |n_{B},n_{F} \rangle & = & 
|n_{B},n_{F}+1 \rangle .  
\end{eqnarray}

With the identifications $n_{B} = \nu$ and $n_{F} = 
\frac{1}{2} (1-\sigma)$, a Landau level may be defined 
as a set $\{ |\eta,n_{B},n_{F}\rangle \}_{\eta}$ of extended 
Fock states. Thus the lowest Landau level corresponds to the 
set $\{ |\eta,0,0\rangle \}_{\eta}$, in which the number of 
bosons and fermions both vanish. The states in this set are 
annihilated by the supercharge $Q$ and its adjoint $Q^{\dagger}$ 
(unbroken supersymmetry), and are therefore associated with 
zero energy. Higher levels $n_{B}>0$ are obtained by repeated 
action of $a_{AC}^{\dagger}$. The supersymmetric partners 
correspond to the sets of Fock states $\{ \{ |\eta, n_{B},1 
\rangle \}_{\eta} \}_{n_{B}}$ that contain one fermion. 
Each such set is obtained from the set $\{ |\eta,n_{B}+1,0\rangle 
\}_{\eta}$ under action of $Q$.   

\section{Conclusions}
An atomic analogue of the Landau quantization based on the 
Aharonov-Casher (AC) effect has been discussed. The effect 
is intimately related to the AC duality between charge and 
magnetic moment. We have shown how symmetric gauge and Landau 
gauge can be realized using two differently shaped distributions 
of charge. Finally, we have argued that supersymmetry plays 
a role in that it makes the zero-point energy to vanish in 
the AC analogue of the Landau effect. Further extension of 
the Landau quantization to other multipole moments should be of 
interest, the most important of which would be that of an 
electric dipole moving in a magnetic field. 

This work could be of interest for creating an atomic quantum 
Hall system that may be realized in a Bose-Einstein condensate. 
In the strong interaction regime of such a system, the present 
result may  provide a realization of the fractional quantum Hall 
effect using electric fields. 

It should be kept in mind that a significant AC Landau quantization 
requires extreme conditions. In order to achieve reasonable 
separation of Landau energies and sufficiently small magnetic 
length $l_{AC}$ a dense charge distribution is needed. Thus 
it will be important to distinguish the present effect from 
those associated with the polarization of the atoms induced 
by this charge. Technical difficulties of this kind must be 
overcome before the effect can be studied in the laboratory. 
\vskip 0.5 cm
We wish to thank Mauritz Andersson for discussions and 
helpful suggestions. E.S.~acknowledges financial 
support from the Swedish Natural Science Research Council (NFR).

\end{document}